\documentclass[reprint,amsmath,amssymb,aps]{revtex4-2}
\usepackage{graphicx}
\usepackage{dcolumn}
\usepackage{bm}
\usepackage{subfig}
\usepackage[justification=justified]{caption}
\usepackage{caption} 
\usepackage{ragged2e}
\usepackage{svg}
\usepackage[colorlinks=true, linkcolor=blue, citecolor=blue, urlcolor=blue]{hyperref}
\usepackage{subcaption}
\usepackage{graphicx}
\usepackage{subfig}
\usepackage{ragged2e}
\usepackage{graphicx, overpic}
\usepackage{float}

\begin{document}
\preprint{APS/123-QED}
\title{Ultra-narrow homogeneous linewidths of erbium-doped silica glass fibers at millikelvin temperatures: magnetic field and temperature dependence}

\author{Farhad Rasekh$^{1,2}$}
\email{farhad.rasekh@ucalgary.ca}
\author{Nasser Gohari Kamel$^{1,2}$}
\author{Mahdi Bornadel$^{1,2}$}
\author{Sourabh Kumar$^{1,2}$}
\author{Erhan Saglamyurek$^3$}
\author{Christoph Simon$^{1,2}$}
\email{christoph.simon@ucalgary.ca}
\author{Daniel Oblak$^{1,2}$}
\email{doblak@ucalgary.ca}
\affiliation{Institute for Quantum Science and Technology, University of Calgary, Calgary, AB, Canada$^1$}
\affiliation{Department of Physics and Astronomy, University of Calgary, Calgary, AB, Canada$^2$}
\affiliation{Lawrence Berkeley National Laboratory, University of California, Berkeley
Berkeley, CA, USA$^3$}

\begin{abstract}
Erbium-doped solids are promising candidates for fiber-based quantum networks due to their emission wavelength, which aligns with the telecom band over which optical fibers exhibit minimal loss. Among these, erbium-doped silica fiber (EDF) stands out for its availability, ease of use, and seamless integration with existing fiber-optic infrastructure. In this work, using the two-pulse photon-echo (2PPE) technique, we measured the homogeneous linewidth of the \( ^4I_{13/2}~\leftrightarrow~^4I_{15/2} \) optical transition under varying magnetic fields and temperatures. We observed an effective homogeneous linewidth of approximately 8~kHz at $\sim 7 \, \mathrm{mK}$ at an optimal magnetic field of 0.09 T, representing over two orders of magnitude improvement compared to earlier reports measured at $T \approx 700\,\mathrm{mK}$. We also present a comprehensive model for the combined magnetic field and temperature dependency of the effective homogeneous linewidth. Additionally, we employed three-pulse photon-echo (3PPE) measurements to investigate spectral diffusion and decoherence processes, and conclude that Two-Level System (TLS) effects are significantly suppressed at sufficiently low temperatures, below $\sim 100 \, \mathrm{mK}$.  

\end{abstract}

\maketitle

\section{Introduction\\}

The ability to store and manipulate quantum information is fundamental for advancing quantum communication \cite{lvovsky2009optical, heshami2016quantum, tittel2025quantum} and other emerging quantum technologies. Rare-earth ions, particularly erbium (Er), offer unique properties that make them highly suitable for such applications\cite{stevenson2022erbium, goldner2015rare}. Rare-earth elements possess a distinctive electronic configuration in which their unfilled 4f electron shell is tightly bound to the atomic nucleus and remains largely shielded from the surrounding environment. As a result, when rare-earth ions are embedded into solid-state hosts, such as crystals or glasses, the 4f electrons retain their intrinsic properties with minimal influence from the host material \cite{liu2006spectroscopic, goldner2015rare}.

Erbium features optical transitions between \( ^4I_{15/2} \) and \( ^4I_{13/2} \)) levels within the 4f-4f configuration, which result in exceptionally narrow homogeneous linewidths at cryogenic temperatures \cite{gritsch2022narrow, fukumori2019optical, zhang2023optical, bottger2024optical, berkman2021sub}. When incorporated into silica fibers, Er ions offer a solid-state platform with distinct advantages \cite{veissier2016optical, saglamyurek2015efficient, jin2015telecom, saglamyurek2015quantum, bornadel2025hole}. Erbium-doped silica fiber is widely used in telecommunications and is compatible with existing fiber-optic infrastructure. Its optical transitions occur at a wavelength of approximately 1.5~\textmu m, within the telecom band, in which optical fiber loss is minimized. This makes Er-doped silica fibers particularly well-suited for long-distance transmission of photons.

Another important property of Er-doped silica fibers is the large inhomogeneous broadening ($\sim 1 \, \mathrm{THz}$) caused by the disordered structure of silica, while the homogeneous linewidths remain narrow. This significant disparity between inhomogeneous and homogeneous linewidths enables spectral tailoring of the absorption profile, which is critical for broadband and multimode storage of optical signals \cite{afzelius2009multimode, wei2024quantum}. Furthermore, under non-zero magnetic fields, erbium ions exhibit spin transitions in the microwave domain, which are compatible with superconducting qubits. This opens up possibilities for microwave-to-optical photon transduction, a key component in hybrid quantum systems \cite{xie2022chip, welinski2019electron}.

Photon echo \cite{kurnit1967photon} techniques, such as two-pulse photon echo (2PPE) and three-pulse photon echo (3PPE), are widely used to study optical coherence and spectral diffusion in solid-state systems. These methods distinguish between homogeneous and inhomogeneous broadening, offering precise insights into the interactions dynamically affecting coherence properties. In Er-doped on silica fiber (EDF), photon echo experiments under controlled temperature and magnetic field conditions enable detailed investigation of the factors influencing homogeneous linewidth \cite{veissier2016optical}. This approach is essential for understanding and improving EDF performance in quantum communication and memory technologies.

To investigate the properties of Er$^{3+}$ in EDF, we performed experiments at millikelvin temperatures and magnetic fields between $0$ and $2 \, \mathrm{T}$. From 2PPE measurements, we observed an effective homogeneous linewidth ($\Gamma_{\text{eff}}$) of approximately $8 \, \mathrm{kHz}$ at $\sim 7 \, \mathrm{mK}$ and $0.09 \, \mathrm{T}$, nearly two orders of magnitude narrower than values previously reported at $\sim 700 \, \mathrm{mK}$~\cite{veissier2016optical}. At around $100 \, \mathrm{mK}$, we identified a change in the temperature dependence of the $\Gamma_{\text{eff}}$, suggesting that the coupled TLS model may become relevant below this temperature~\cite{Macfarlane2006elastic, hegarty1983photon, ding2016multidimensional, ding2020probing}. Our observations lead us to modify a previous model of the $\Gamma_{\text{eff}}$ to accurately provide a coherent description of its magnetic field and temperature dependency at low temperature regime. Complementary 3PPE measurements revealed contributions from spin-spin interactions and two-level systems (TLS) to long-timescale decoherence in EDF, with TLS-related dephasing reduced by more than an order of magnitude from $\sim 425 \, \mathrm{mK}$ to $\sim 7 \, \mathrm{mK}$. Finally, the branching ratio to the other Zeeman level displayed a non-monotonic magnetic-field dependence, with enhanced values near $0.09 \, \mathrm{T}$ that point to a crossover between hyperfine- and Zeeman-dominated regimes.

The paper is organized as follows. Section~\ref{sec:setup} describes the sample, experimental setup, and photon echo measurement techniques. Section~\ref{sec:results} presents the results and analysis of the effective homogeneous linewidth, including its dependence on magnetic field and temperature, as well as spectral diffusion across different regimes. By examining the system across different regimes, we are able to identify the conditions that yield the optimal homogeneous linewidth. Finally, Section~\ref{ssec:discussion_outlook} provides a discussion and outlook.

\section{Methods and Sample\\}\label{sec:setup}

\begin{figure*}[t]
\centering
\includegraphics[width=\textwidth,height=0.35\textheight,keepaspectratio]{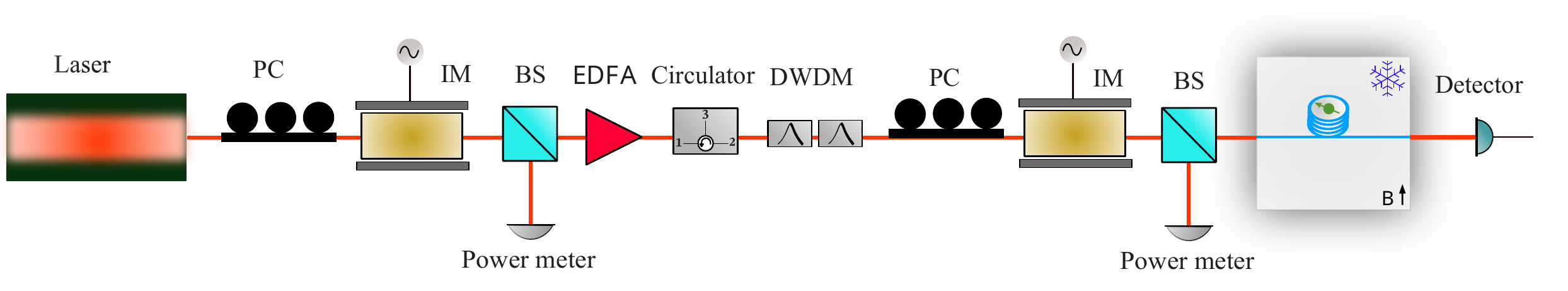}
\caption{\justifying Schematic of the experimental setup for photon echo measurements in erbium-doped silica fiber (EDF). A continuous-wave 
diode laser at 1532.5 nm is pulsed using an intensity modulator (IM), with polarization controlled by a polarization controller (PC). The output is monitored by a beam splitter (BS) and a power meter. The light is amplified by an erbium-doped fiber amplifier (EDFA), with a circulator preventing back reflections. Dense Wavelength-Division Multiplexing (DWDM) filters suppress amplified spontaneous emission before the light enters the EDF housed in a dilution refrigerator. The photon echo signal is detected by a fast photodetector.}
\label{sec1-fig:1}
\end{figure*}

Photon-echo techniques provide a means to measure the coherence properties of atomic ensembles by inducing and detecting rephasing events. In the 2PPE sequence, a pair of laser pulses are applied to excite a subset of erbium ions into a superposition state between the ground state (\( ^4I_{15/2} \)) and the excited state (\( ^4I_{13/2} \)). The first pulse, ideally a \( \frac{\pi}{2} \)-pulse, initiates the superposition, after which the ions begin to dephase due to variations in their resonant frequencies caused by inhomogeneous broadening. After a time delay \( t_{12} \), a \( \pi \)-pulse is applied to reverse the dephasing, causing the ions to rephase and emit a photon-echo at \( 2t_{12} \). The 3PPE extends this sequence by adding a third pulse after \( t_{12} \), separated by a time interval \( t_{23} \). This additional pulse induces a second rephasing process, allowing for the investigation of spectral diffusion by varying either \( t_{12} \) or \( t_{23} \) while keeping the other fixed.

In our experiment, we used 10 meters of EDF with a doping concentration of 200 ppm (INO S/N 402-28254), spooled in layers around a 4 cm diameter copper cylinder to fit inside a BLUEFORS dilution refrigerator. To enhance uniform cooling, aluminum foil and cryogenic heat-conducting paste were applied between the fiber layers. The refrigerator provided temperatures as low as $\sim 7 \, \mathrm{mK}$. The setup was positioned at the center of a superconducting magnet capable of generating magnetic fields up to $B = 2 \, \mathrm{T}$ in one direction along the axis of the fiber spool. The sample was cooled to $\sim 7 \, \mathrm{mK}$, resulting in an optical depth of $\alpha L \approx 3$ at $\lambda = 1532.5 , \mathrm{nm}$. These EDFs also contain co-dopant ions, such as $^{27}$Al, which are added to prevent clustering of the Er$^{3+}$ ions in the fiber. Note that all temperatures mentioned in this paper refer to the temperature of the cryostat cold plate, and the actual temperature of the sample may differ slightly.

As shown in Figure~\ref{sec1-fig:1}, two intensity modulators (IMs) gated the continuous-wave (CW) output of our laser diode at 1532.5 nm, which has a nominal linewidth of less than 1 MHz, to generate the optical pulses required for the 2PPE and 3PPE experiments. The IMs were controlled using an arbitrary waveform generator (Tektronix AWG70002A) and a function generator (Tektronix AFG3102). Power meters positioned after the beam splitters (BS) monitored the IM output, allowing optimization of the applied DC voltage. Typical pulse widths were 4~ns for $\frac{\pi}{2}$-pulses and 8~ns for $\pi$-pulses. A circulator acted as an optical isolator to prevent reflected light from re-entering the setup. An erbium-doped fiber amplifier (EDFA) provided a peak power of approximately 100~mW at the sample, and two 50~GHz Dense Wavelength-Division Multiplexing (DWDM) filters suppressed amplified spontaneous emission from the EDFA. Photon echo signals were detected using a fast photodiode (New Focus 1554-B) and recorded with a digital oscilloscope. The time between measurements was carefully selected to prevent persistent spectral holes \cite{bornadel2025hole} in the sample.

\begin{figure*}[htb]
  \centering
  \subfloat[]{%
    \includegraphics[width=0.49\textwidth]{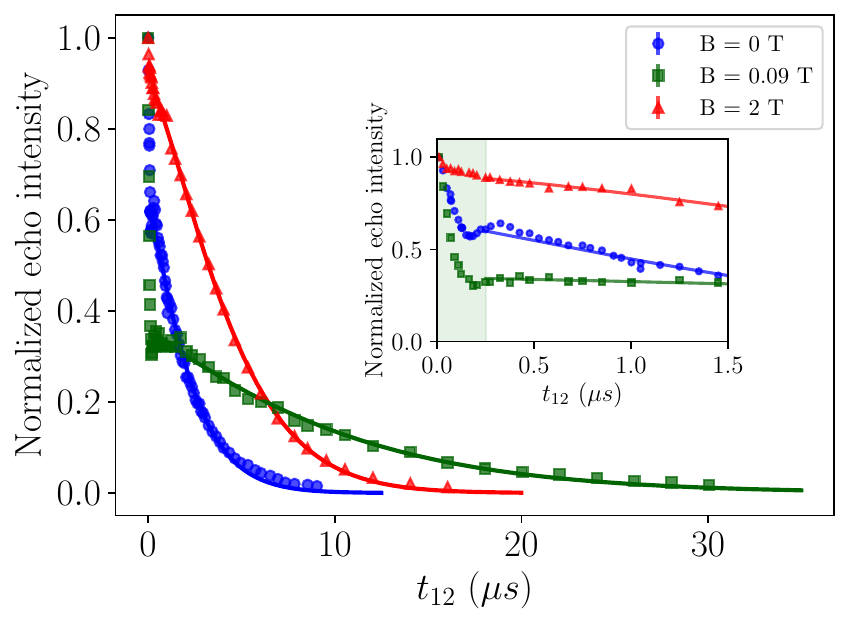}%
    \label{fig:2PPE}
  }\hfill
  \subfloat[]{%
    \includegraphics[width=0.49\textwidth]{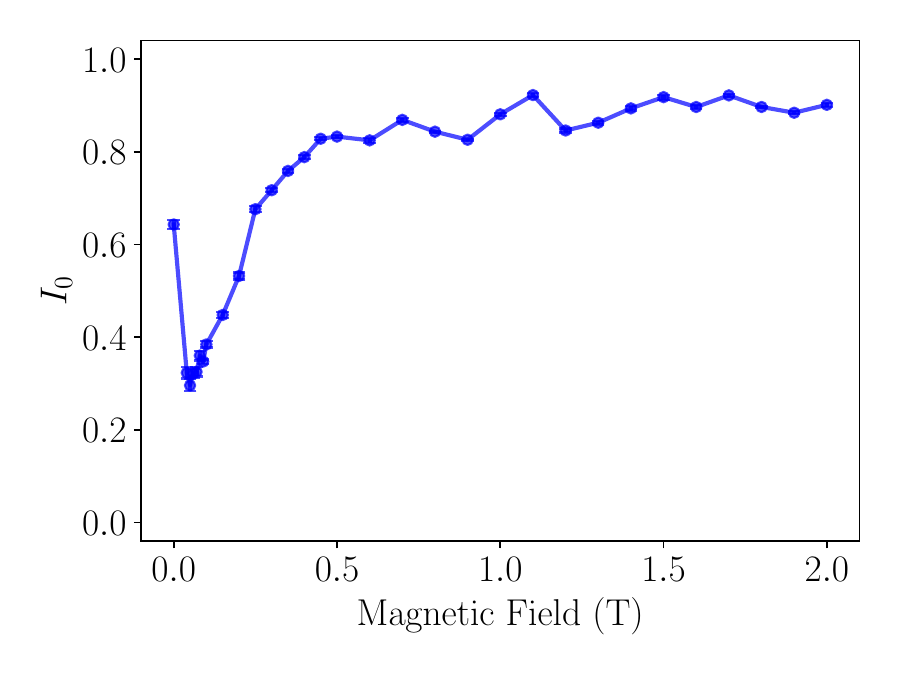}%
    \label{fig:homo}
  }


  \subfloat[]{%
    \includegraphics[width=0.49\textwidth]{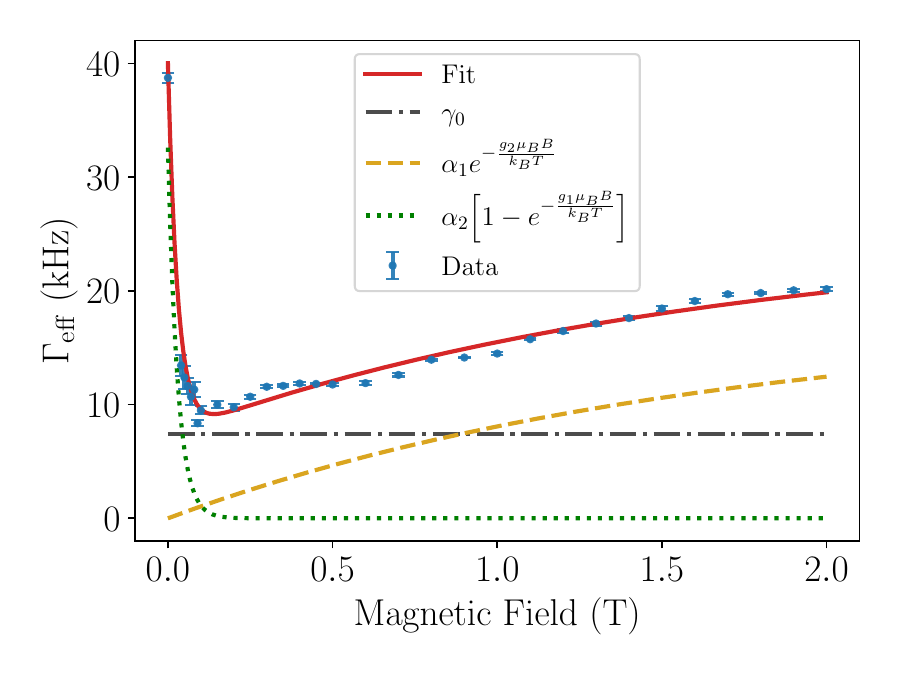}%
    \label{fig:Field_dependence}
  }\hfill
  \subfloat[]{%
    \includegraphics[width=0.49\textwidth]{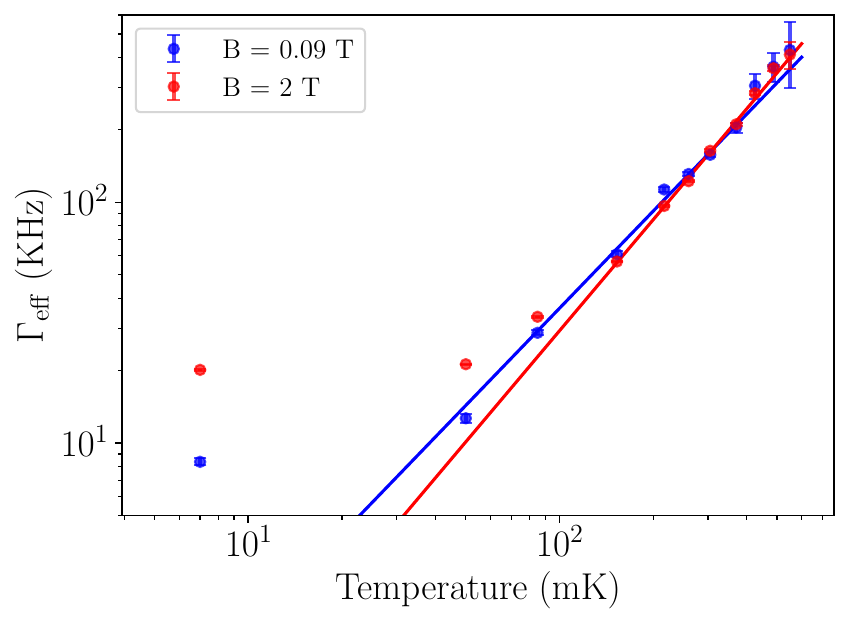}%
    \label{fig:Temp_dependence}
  }

  \caption{%
    \justifying
    (a) Two-pulse photon echo (2PPE) decay measured at $T \approx 7\,\mathrm{mK}$ under magnetic fields of $B = 0 \, \mathrm{T}$ (blue circles), $0.09 \, \mathrm{T}$ (green squares), and $2 \, \mathrm{T}$ (red triangles). The plot depicts the echo intensity amplitude as a function of the delay time between the $\pi$- and $\pi/2$-pulses in the echo sequence. The dots represent experimental data points, fitted by Mims’ law (equation \ref{eq:1.2})), and the inset plot is a zoom of the first 1.5 µs. (b) Magnetic field dependence of the zero-delay echo intensity ($I_0$ in equation \ref{eq:1.2}) following modulation in the echo signal. The line is included to guide the eye and is not a fit. (c) Magnetic-field dependence of the effective homogeneous linewidth ($\Gamma_{\text{eff}} = 1/\pi T_M$, see Eq.~(\ref{eq:1.2})) at $T \approx 7\,\mathrm{mK}$. The linewidth decreases with increasing field before rising again at higher fields, indicating an optimal field where $\Gamma_{\text{eff}}$ is minimized. The data are well described by Eq.~\ref{eq:1.3}, and the contributions of the individual terms are also shown in the figure. (d) Temperature dependence of the effective linewidth at B = 0.09 T (blue) and B = 2 T (red). For temperatures above $\sim$100 mK, this dependence follows $\alpha T^n$, with $n = 1.34 \pm 0.04$ and $n = 1.53 \pm 0.01$ for 0.09 T and 2 T, respectively. At very low temperatures, below $\sim$100 mK, the linewidth approaches a constant value, which can be explained by coupled TLSs as discussed in the main text.}
  \label{fig:dephasing_panels}
\end{figure*}

In 2PPE, 4~ns and 8~ns optical pulses separated by \( t_{12} \) were applied, with the photon echo detected at \( 2t_{12} \). In 3PPE, three 4~ns pulses were used, separated by \( t_{12} \) and \( t_{23} \), with the echo observed at \( t_{23} + 2t_{12} \). By varying \( t_{12} \) in 2PPE, the optical coherence time of the system is determined. In 3PPE, varying either \( t_{12} \) or \( t_{23} \) while keeping the other fixed allows the measurement of spectral diffusion and the identification of dynamic decoherence mechanisms within the sample.

\section{Results\\}\label{sec:results}

In this section, we present the 2PPE and 3PPE experiments for different temperatures ($\sim 7$–$550 \, \mathrm{mK}$) and magnetic fields ($0$–$2 , \mathrm{T}$). Subsection \ref{sectionA} presents the results of 2PPE measurements, in which \ref{sectionA1} shows the effective homogeneous linewidth at $T \approx 7 \, \mathrm{mK}$ and $B = 2 \, \mathrm{T}$, and \ref{sectionA2} discusses the temperature and magnetic field dependence of the effective homogeneous linewidth. Finally, subsection \ref{sectionC} focuses on investigating spectral diffusion and decoherence mechanisms using 3PPE experiments.

 \subsection{2PPE measurements}\label{sectionA}
We conducted this 2PPE experiment to extract the coherence properties of Er ions in EDF. Three examples of echo intensity as a function of the pulse separation for three different magnetic fields (0, 0.09, and 2~T) at a temperature of $\sim 7 \, \mathrm{mK}$ are shown in Fig.~\ref{fig:2PPE}. These three specific magnetic fields were selected for the following reasons: at 0 T, the spin-spin interactions is expected to be the major source of decoherence and spectral diffusion; at 0.09 T, the narrowest homogeneous linewidth ($\sim$8 kHz) was observed; and finally, the highest magnetic field used in the experiment, 2 T, was also examined.

\subsubsection{Effective Homogeneous Linewidth}\label{sectionA1}
We observed a stretched exponential for echo intensity vs $t_{12}$, indicating the presence of spectral diffusion, which is fitted with Mims' expression:

\begin{equation}
I(t_{12}) = I_0 e^{-2 (\frac{2 t_{12}}{T_M})^x}
\label{eq:1.2}
\end{equation}

where $I_0$ is the zero-delay echo intensity (i.e., extrapolated at $t_{12} = 0$), $\Gamma_{\text{eff}} = \frac{1}{\pi T_M}$ represents the full width at half maximum (FWHM) of the effective homogeneous linewidth, $x$ is the Mim's exponent, which indicates how stretched the exponential function is.

An initial drop accompanied by sinusoidal modulation, shown in the inset plot of Fig.~\ref{fig:2PPE}, arises from the super-hyperfine coupling between the nuclear spins of Er ions and co-dopants such as $^{27}$Al in the silica matrix \cite{wannemacher1991nuclear,staudt2006investigations}. As observed, this modulation is more pronounced at lower magnetic fields. This occurs because, at these fields, the echo pulse sequence simultaneously excites transitions from multiple super-hyperfine levels, leading to coherent interference that produces the echo modulation, as previously reported \cite{kim1989hyperfine,kim1991hyperfine,macfarlane2007optical,mitsunaga1992stimulated,whittaker1982hyperfine}. We fitted all our 2PPE data sets with Mims' law after $t = 250 \, \mathrm{ns}$, when the initial drop or modulation is largely diminished.. A more detailed investigation of this modulation, and its comparison with the model proposed in \cite{mitsunaga1992stimulated}, would require stronger and more numerous oscillations. Additional experiments, such as spectral hole burning, could also be performed to observe side holes at the same sinusoidal frequency, indicating coupling to nuclear spins in the environment, as reported in \cite{staudt2006investigations}. These investigations, however, are the main objective of this paper.

As shown, the echo intensity decreases following the modulation, with the extent of the drop depending on the magnetic field. This drop reduces the number of ions that act coherently and, thus, will limit any process, which relies on coherent evolution of the ensemble, such as quantum memory. In Fig.~\ref{fig:homo}, the zero-delay echo intensity $I_0$ from Eq.~(\ref{eq:1.2}), obtained from the normalized dataset, is plotted as a function of the magnetic field.
At lower magnetic fields, at which the effective linewidth is narrower (Fig.~\ref{fig:Field_dependence}), $I_0 \sim$ 0.3. In contrast, at higher magnetic fields, the linewidth increases by about a factor of two (Fig.~\ref{fig:Field_dependence}), while the $I_0$ approaches unity. Therefore by increasing the magnetic field from $\sim$0.09 to $\sim$0.4T and consequently the $\Gamma_{\text{eff}}$ from $\sim$8 to $\sim$12 kHz, the zero-delay echo intensity, $I_0$ in Eq.~(\ref{eq:1.2}), increases from $\sim$30 to $\sim$80\%.

As shown in Fig.\ref{fig:Field_dependence}, at $T \approx 7 \, \mathrm{mK}$ and $B = 0.09 \, \mathrm{T}$, we observed a homogeneous linewidth of about $\sim 8~\mathrm{kHz}$ ($\sim 40~\mu\mathrm{s}$ coherence time), which is over two orders of magnitude narrower than previously reported \cite{veissier2016optical}. The dependence of coherence times on magnetic field and temperature is explored and discussed in the next section.


\subsubsection{Magnetic field and temperature dependency of effective homogeneous linewidth $\Gamma_{\text{eff}}$}\label{sectionA2}

In EDF, decoherence can occur through a range of interactions. One such
interaction, is between the probed Er$^{3+}$ ions and pairs of coupled Er$^{3+}$ spins in the surrounding lattice, which undergo random spin-state exchanges (Er$^{3+}$–Er$^{3+}$ spin flip-flops). In our EDF sample, where the Er$^{3+}$ ions are embedded in an amorphous host, decoherence can also arise from TLSs associated with structural disorder. These TLSs, formed by local potential minima that allow quantum tunneling, contribute to decoherence \cite{phillips1972tunneling}.  

Moreover, as discussed in \cite{Macfarlane2006elastic}, Er$^{3+}$ ions in amorphous materials, such as EDF, exhibit a highly anisotropic $g$-factor, which largely suppresses direct Er$^{3+}$–Er$^{3+}$ flip-flop interactions. Although this anisotropy suppresses flip-flops more strongly than in crystals, the measured variation of $\Gamma_{\text{eff}}$ with magnetic field is actually larger in EDF. In crystalline systems, for which such interactions are more common, high magnetic fields can freeze them out more effectively. The unexpectedly stronger magnetic-field dependence in EDF therefore points to an additional decoherence mechanism, namely interactions with TLSs. In addition to elastic TLSs (nonmagnetic), there exists a second class known as spin-elastic TLSs, which possess a magnetic character \cite{Macfarlane2006elastic}. Elastic TLSs affect coherence without any dependence on the applied magnetic field, while spin-elastic TLSs exhibit magnetic behavior and can normally be suppressed at high fields. As a result, stronger magnetic fields reduce spin-flip processes among TLSs, thereby decreasing their contribution to $\Gamma_{\text{eff}}$.  

Experimentally, we obtain the magnetic field dependence of $\Gamma_{\text{eff}}$ in our EDF sample at $T \approx 7 \, \mathrm{mK}$ shown in Fig~\ref{fig:Field_dependence}. The data reveals a minimum in $\Gamma_{\text{eff}}$ at low fields, followed by an increase, which appears to saturate at higher fields. We find that this dependence can be fitted to the formula
\begin{align}
\Gamma_{\mathrm{eff}}(B) = \gamma_0 + \alpha_1\exp\left(-\frac{g_1 \mu_B B}{k_B T}\right) +  \nonumber\\
\alpha_2 \left(1 - \exp\left(-\frac{g_2 \mu_B B}{k_B T}\right)\right) \ ,
\label{eq:1.3}
\end{align}
where $\mu_B$ is the Bohr magneton, $k_B$ is the Boltzmann constant, $B$ is the applied external magnetic field, and $T \approx 7 \, \mathrm{mK}$. The parameters $g_i$ (for $i = 1,2$) represent the effective $g$-factors for the coupled Er–magnetic TLS systems in the glass, while $\alpha_i$ (for $i = 1,2$) describe the coupling strengths of the Er ions to different decoherence sources modeled as TLS interactions. Fitting the data with this model yields the parameters  
\( \Gamma_0 = 7.42 \pm 0.14 \, \mathrm{kHz} \),  
\( \alpha_1 = 32.60 \pm 0.32 \, \mathrm{kHz} \),  
\( \alpha_2 = 17.62 \pm 0.49 \, \mathrm{kHz} \),  
\( g_1 = 0.3507 \pm 0.0092 \), and  
\( g_2 = 0.0064 \pm 0.0004 \).  

Building on the initial discussion on relevant decoherence sources, we ascribe the magnetic-field independent first, $\gamma_0$, term of Eq.~(\ref{eq:1.3}) (black dash-dot line in Fig.~\ref{fig:Field_dependence}), to contributions from the single-ion homogeneous linewidth, excitation-induced decoherence such as instantaneous spectral diffusion (ISD) \cite{thiel2014measuring}, and elastic TLSs. It defines the lower bound of $\Gamma_{\text{eff}}$ as a function of magnetic field, such that at zero field the linewidth is $\Gamma_{\mathrm{eff}}(0) = \gamma_0 + \alpha_1$.  

The second term (orange dashed line in Fig.~\ref{fig:Field_dependence}) represents the spin-elastic TLS contribution to $\Gamma_{\text{eff}}$, with $\alpha_1$ and $g_1$ corresponding to the coupling strength and effective $g$-factor, respectively. However, this term alone does not fully capture the observed field dependence.  

The third term (green dotted line in Fig.~\ref{fig:Field_dependence}) in Eq.~(\ref{eq:1.3}) captures an additional decoherence mechanism that becomes significant at higher magnetic fields, as revealed by the data at $T \approx 7 \, \mathrm{mK}$ in Fig.~\ref{fig:Field_dependence}. The model and fitted values suggest that the second and third terms may correspond to distinct subsets of Er$^{3+}$ ions, each interacting with a different class of TLS environment. The class giving rise to the third term is unusual, as its contribution to decoherence increases with magnetic field, which could be due to coupled TLSs, as discussed in Section~\ref{ssec:discussion_outlook}.
Although this interpretation is still hypothetical, it is consistent with the distinct field dependencies observed. Notably, the third term imposes a saturation limit on $\Gamma_{\text{eff}}$ at high fields. 

The temperature dependence of $\Gamma_{\text{eff}}$ is presented in Fig.~~\ref{fig:Temp_dependence}, comparing results at magnetic fields of 0.09 T and 2 T. The data were fitted with a power-law dependence, $\Gamma_{\text{eff}}(T) \propto \alpha T^n$, yielding exponents $n = 1.34 \pm 0.04$ for 0.09 T and $n = 1.53 \pm 0.01$ for 2 T. These values are consistent with previous studies~\cite{Macfarlane2006elastic, hegarty1983photon, ding2016multidimensional, ding2020probing}. Below approximately 100 mK, $\Gamma_{\text{eff}}(T)$ saturates to a constant value. A more detailed discussion of the magnetic field and temperature dependence of $\Gamma_{\text{eff}}$ is provided in Section~\ref{ssec:discussion_outlook}.

\subsection{3PPE measurements}\label{sectionC}

Spectral diffusion in EDF, referring to the time-dependent broadening of optical transitions due to slow fluctuations in the local environment, was investigated using 3PPE measurements. In these experiments, two initial pulses separated by a time delay \( t_{12} \) establish a spectral grating in the ensemble, and a third pulse, arriving after a waiting time \( t_{23} \), probes the evolution of this grating. Changes in the echo intensity as a function of \( t_{23} \) reflect time-dependent broadening of the homogeneous linewidth.

Measurements were conducted at temperatures ranging from $\sim 7 \, \mathrm{mK}$ to $425 \, \mathrm{mK}$ under three magnetic field conditions: $0$, $0.09$, and $2 \, \mathrm{T}$. Two types of 3PPE sequences were employed: in one, \( t_{12} \) was fixed while the echo intensity was recorded as a function of \( t_{23} \); in the other, \( t_{23} \) was held constant and the echo amplitude was measured versus \( t_{12} \). This dual approach enabled characterization of both the intrinsic coherence decay and the evolution of spectral diffusion over time.

In the idealized two-level system, the intensity of the stimulated photon echo \( I(t_{12}, t_{23}) \) decays according to \cite{bottger2006optical}

\begin{equation}
I(t_{12}, t_{23}) = I_0 e^{-2t_{23}/T_1} e^{-4\pi t_{12} \Gamma_{\mathrm{eff}}(t_{12}, t_{23})},
\end{equation}

where \( I_0 \) is the maximum echo intensity (a scaling coefficient), \( T_1 \) is the excited-state lifetime, and \( \Gamma_{\mathrm{eff}}(t_{12}, t_{23}) \) represents the effective homogeneous linewidth that evolves due to spectral diffusion.

However, in our EDF system, population trapping in the intermediate Zeeman level introduces effective three-level dynamics, which modifies the echo intensity behavior. In this case, a corrected form of the echo decay is used according to~\cite{sinclair2010spectroscopic,sun2012optical}:

\begin{align}
I(t_{12}, t_{23}) &= \nonumber \\
& I_0 \left\{ e^{-t_{23}/T_1} + \frac{\beta}{2} \frac{T_Z}{T_Z - T_1} 
\left( e^{-t_{23}/T_Z} - e^{-t_{23}/T_1} \right) \right\}^2 \nonumber \\
& \times e^{-4\pi t_{12} \Gamma_{\mathrm{eff}}(t_{12}, t_{23})}
\label{eq:3PPE_decay_model}
\end{align}

Where \( T_Z \) is the lifetime of the intermediate Zeeman level and \( \beta \) is the branching ratio which denotes the fraction of spontaneous emission from the excited state that decays into this long-lived Zeeman level. In our analysis, we use \( T_1 = 9 \)~ms, determined by fitting the full model to our echo decay data, which is in the range reported by previous studies \cite{vavrak2024heat,staudt2006investigations,dajczgewand2015optical,veissier2016optical}. The values of \( T_Z \), which are on the order of seconds, are taken from our previous study on Zeeman population storage in the same temperature and magnetic field regime~\cite{bornadel2025hole}.

The time-dependent effective linewidth \( \Gamma_{\mathrm{eff}}(t_{12}, t_{23}) \) captures the contributions from environmental interactions that cause spectral diffusion. The full spectral diffusion model is described by
\begin{equation}
\begin{split}
    \Gamma_{\mathrm{eff}} (t_{12}, t_{23}) = \Gamma_0 + \frac{1}{2} \Gamma_{\mathrm{SD}} \left[ R_{\mathrm{SD}} t_{12} + \left(1 - e^{-R_{\mathrm{SD}} t_{23}} \right) \right] \\
    + \Gamma_{\mathrm{TLS}} \log_{10}\left(\frac{t_{23}}{t_0}\right),
\end{split}
\end{equation}
where \( \Gamma_0 \) is the intrinsic homogeneous linewidth in the absence of spectral diffusion. While \( \Gamma_0 \) is treated as a free parameter in the 3PPE fits, the extracted values are consistent with those obtained from independent 2PPE measurements. The second term models time-dependent broadening due to magnetic dipole-dipole interactions between Er$^{3+}$ ions, which cause their transition frequencies to shift in response to spin flip-flops occurring in the environment. The parameter \( \Gamma_{\mathrm{SD}} \) represents the maximum contribution to the linewidth from this interaction, while \( R_{\mathrm{SD}} \) is the characteristic spin-flip rate of the surrounding perturbing spins. This description follows the Lorentzian diffusion model typically used in electron and nuclear spin resonance to characterize decoherence arising from randomly timed spin transitions~\cite{bottger2006optical}. The third term accounts for coupling to tunneling TLS, which are abundant in amorphous hosts such as silica. This contribution results in logarithmic broadening over time and is parameterized by \( \Gamma_{\mathrm{TLS}} \), with \( t_0 \) denoting a reference timescale, which is the minimum measurement timescale. Here, \( \Gamma_{\mathrm{TLS}} \) reflects slow spectral diffusion 
and represents a long-timescale average effects of all TLS contributions, distinct from the short-timescale effects analyzed in the 2PPE measurements.~\cite{black1977spectral,breinl1984spectral,littau1992dynamics,silbey1996time,koedijk1996spectral}.

The spectral diffusion dynamics may be most directly probed by examining the effective linewidth as \( t_{12} \to 0 \). In our case, \( t_{12} \) is small compared to \( t_{23} \), and the contribution from the linear \( R_{\mathrm{SD}} t_{12} \) term becomes negligible consistent with the echo decay in our 3PPE measurements which showed negligible dependence on \( t_{12} \). The effective linewidth is then approximately described by
\begin{equation}
\Gamma_{\mathrm{eff}}(t_{23}) = \Gamma_0 + \frac{1}{2}\Gamma_{\mathrm{SD}} \left( 1 - e^{-R_\mathrm{SD} t_{23}} \right) + \Gamma_{\mathrm{TLS}} \log_{10}\left( \frac{t_{23}}{t_0} \right),
\label{eq:Gamma_eff}
\end{equation}
where the broadening is dominated by spectral diffusion during the waiting time \( t_{23} \). This simplified form was used to model the echo decay and extract the contributions from different dephasing mechanisms.

\begin{figure*}[htb]
  \centering

  \subfloat[]{%
    \includegraphics[width=0.49\textwidth]{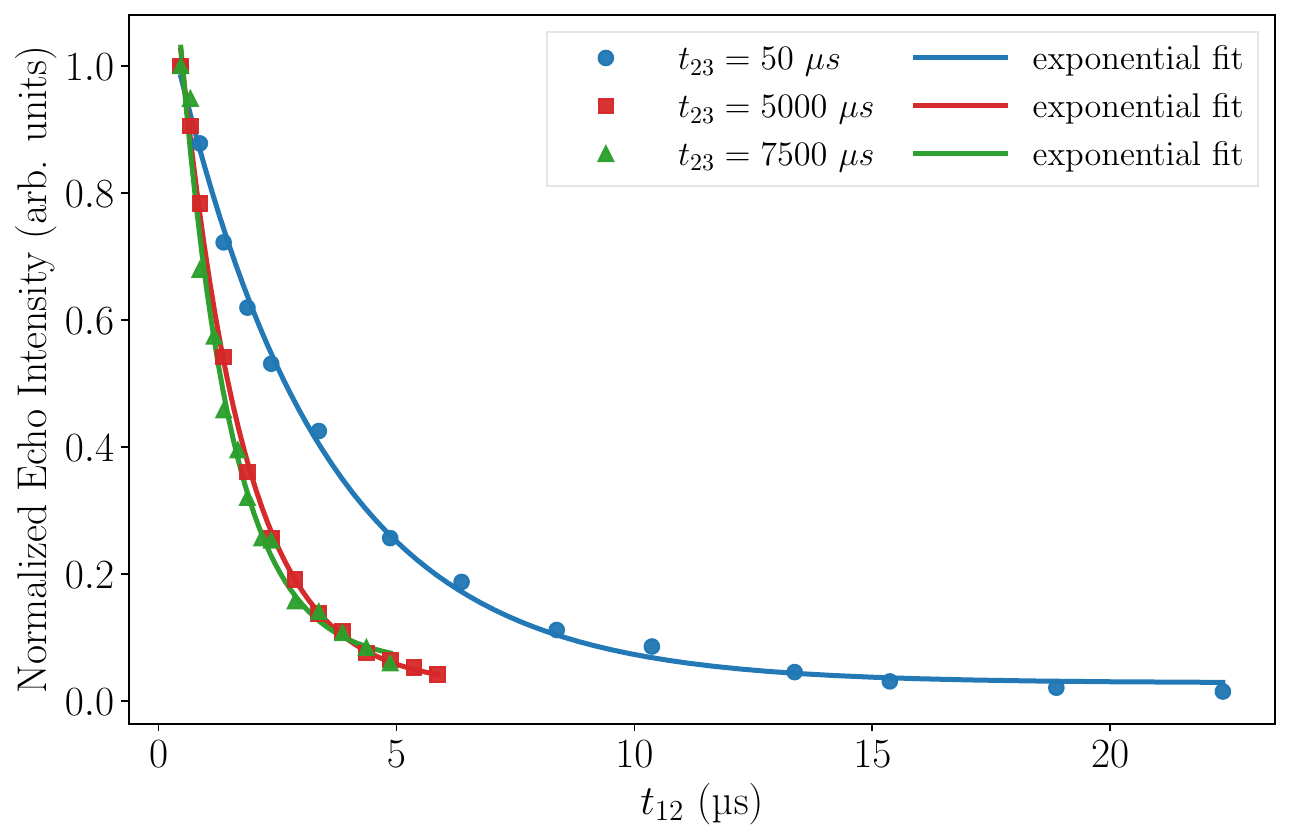}%
    \label{fig:EchoGamma_t12}
  }\hfill
  \subfloat[]{%
    \includegraphics[width=0.49\textwidth]{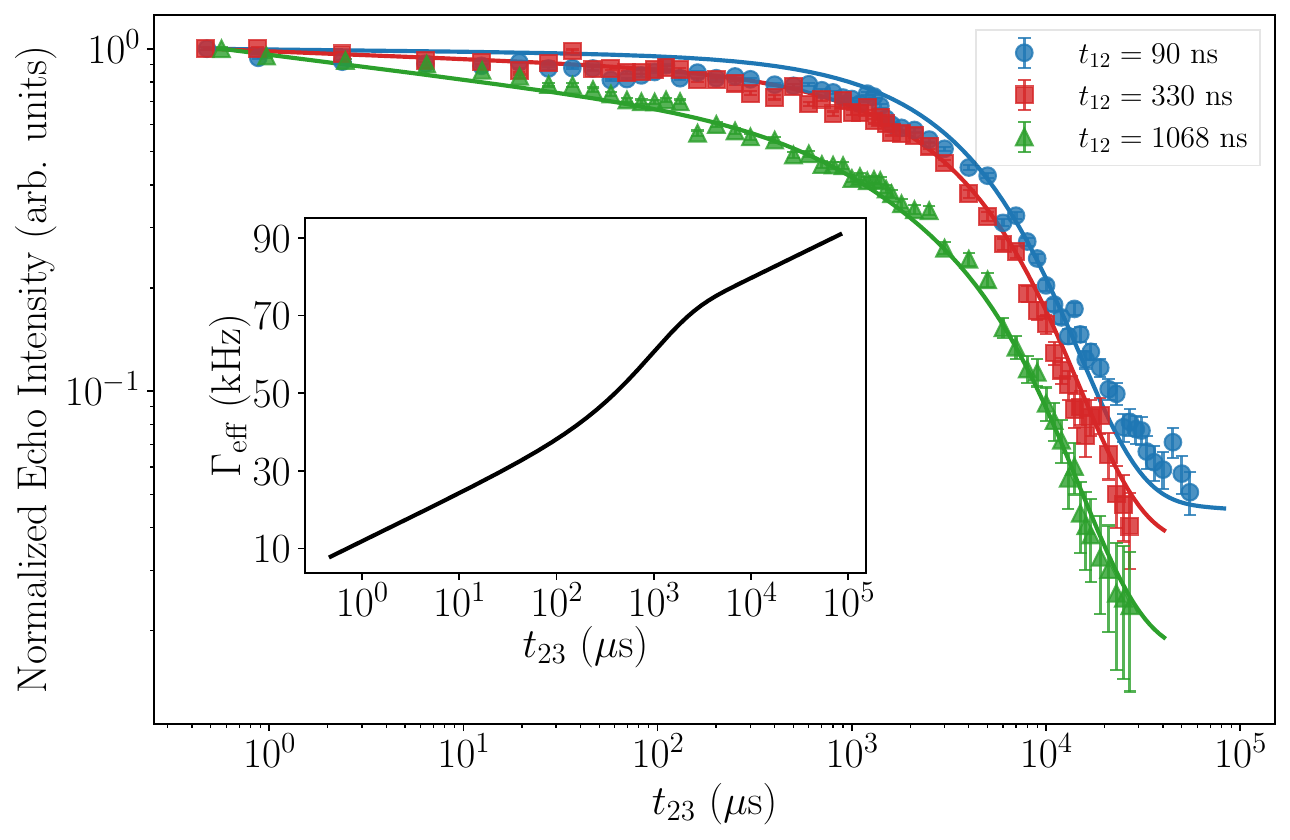}%
    \label{fig:EchoGamma_t23}
  }

  
  \subfloat[]{%
    \includegraphics[width=0.49\textwidth]{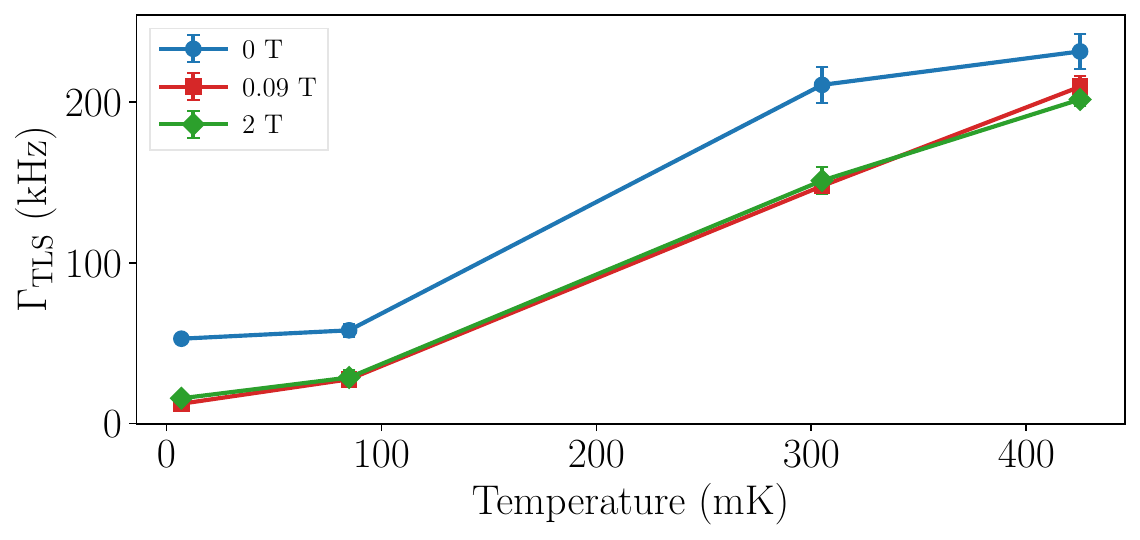}%
    \label{fig:Gamma_TLS}
  }\hfill
  \subfloat[]{%
    \includegraphics[width=0.49\textwidth]{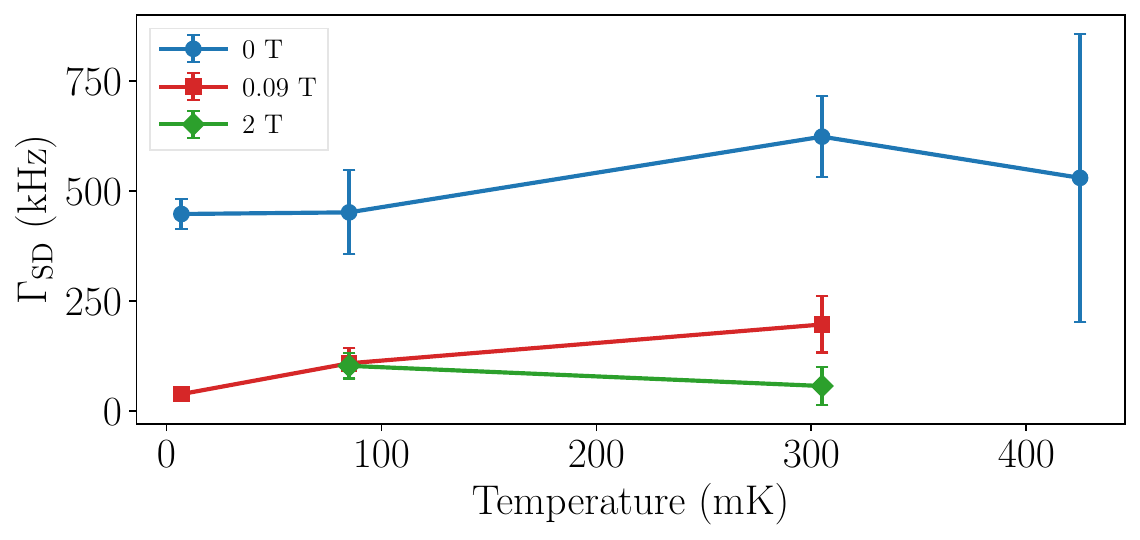}%
    \label{fig:Gamma_SD}
  }


  \subfloat[]{%
    \includegraphics[width=0.49\textwidth]{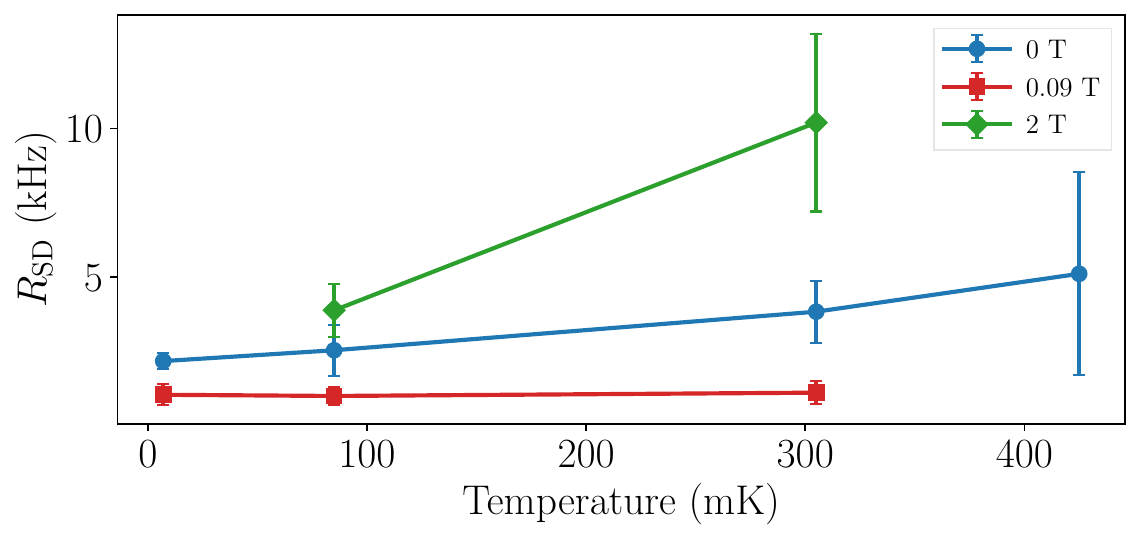}%
    \label{fig:R_SD}
  }\hfill
  \subfloat[]{%
    \includegraphics[width=0.49\textwidth]{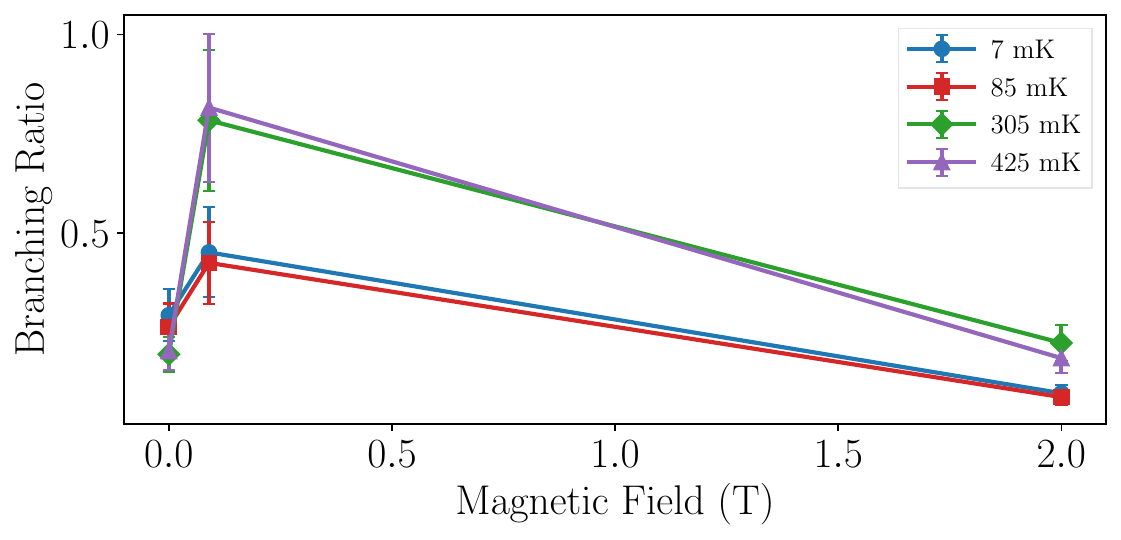}%
    \label{fig:Branching_Ratio}
  }

  \caption{%
    \justifying
    (a) Three-pulse photon echo measurements at $\sim$ 7~mK and 0.09~T with normalized echo intensity plotted versus \( t_{12} \) for three fixed waiting times of \( t_{23} \): \(50~\mu\mathrm{s},\, 5000~\mu\mathrm{s},\) and \(7500~\mu\mathrm{s} \). Solid lines are single-exponential fits \( I \propto \exp[-4\pi t_{12}\,\Gamma_{\mathrm{eff}}(t_{23})] \). (b) Three-pulse photon echo measurements at $\sim$ 7~mK and 0.09~T with normalized echo intensity plotted versus \( t_{23} \) for three fixed values of \( t_{12} \): 90~ns, 330~ns, and 1068~ns. Solid lines represent fits to the three-level echo decay model (Eq.~\ref{eq:3PPE_decay_model}) using the spectral diffusion model (Eq.~\ref{eq:Gamma_eff}) for \( \Gamma_{\mathrm{eff}}(t_{23}) \). The inset shows the predicted values of the effective linewidth \( \Gamma_{\mathrm{eff}} \) derived from the fitted parameters. (c-f) Extracted dephasing parameters from spectral diffusion fits at varying temperatures and magnetic fields. (c) TLS coupling strength \( \Gamma_{\mathrm{TLS}} \) vs temperature. (d) Spectral diffusion amplitude \( \Gamma_{\mathrm{SD}} \) vs temperature. (e) Spectral fluctuation rate \( R_{\mathrm{SD}} \) vs temperature. (f) Branching ratio from the excited state to the ground-state Zeeman level vs magnetic field.
  }
  \label{fig:dephasing_panels}
\end{figure*}

Fig.~\ref{fig:EchoGamma_t12} shows 3PPE measurements at $\sim$ 7~mK and 0.09~T, where the echo intensity is recorded as a function of \( t_{12} \) for three fixed waiting times \( t_{23} = 50~\mu\mathrm{s},\, 5000~\mu\mathrm{s},\) and \(7500~\mu\mathrm{s} \). Each trace is well fit by a single exponential \( I \propto \exp[-4\pi t_{12}\,\Gamma_{\mathrm{eff}}(t_{23})] \), as expected from Eq.~\ref{eq:3PPE_decay_model} and Eq.~\ref{eq:Gamma_eff} for fixed \( t_{23} \).
This behavior indicates that \( \Gamma_{\mathrm{eff}} \) is effectively governed by its \( t_{23} \) dependence over the range used, consistent with the small-\( t_{12} \) limit where the linear \( R_{\mathrm{SD}} t_{12} \) term is negligible.
In a complementary configuration, Fig.~\ref{fig:EchoGamma_t23} presents representative 3PPE measurements at $\sim 7\,\mathrm{mK}$ and $0.09\,\mathrm{T}$, where the echo intensity is recorded as a function of \( t_{23} \) for three fixed values of \( t_{12} \)  = 90~ns, 330~ns, and 1068~ns. The echo decays more rapidly at longer \( t_{12} \), consistent with increased dephasing sensitivity due to spectral diffusion. The fitted curves, based on  echo intensity decay model \ref{eq:3PPE_decay_model} and the full spectral diffusion expression \ref{eq:Gamma_eff} for \( \Gamma_{\mathrm{eff}} \), follow the experimental data across the entire delay range. The inset shows the expected effective linewidth evolution over time, calculated using the fitted parameters: \( \Gamma_0 = 7.96~\pm~0.48 \)~kHz, \( \Gamma_{\mathrm{TLS}} = 12.24~\pm~0.90 \)~kHz, \( \Gamma_{\mathrm{SD}} = 37.77~\pm~4.18 \)~kHz, and \( R_{\mathrm{SD}} = 1.02~\pm~0.25 \)~kHz. At early delays, the logarithmic increase reflects dominant coupling to tunneling TLSs, while at later times, the exponential saturation indicates increasing influence from spin flip-flop dynamics. This confirms that both fast and slow decoherence channels are active in the EDF system under these conditions.

Fig.~\ref{fig:Gamma_TLS}, \ref{fig:Gamma_SD}, and \ref{fig:R_SD} summarizes the fitted dephasing parameters across different conditions, showing the temperature dependence of the TLS contribution \( \Gamma_{\mathrm{TLS}} \), the spectral diffusion amplitude \( \Gamma_{\mathrm{SD}} \), and the characteristic spin-flip rate \( R_{\mathrm{SD}} \), respectively. Fig.~\ref{fig:Branching_Ratio} presents the branching ratio from the excited state to the intermediate Zeeman level as a function of magnetic field.

As seen in Fig.~\ref{fig:Gamma_TLS}, \( \Gamma_{\mathrm{TLS}} \) increases sharply with temperature, indicating strong coupling to tunneling TLSs in the amorphous silica host. At higher temperatures, this contribution becomes the primary source of dephasing sensitivity in the fit, limiting our ability to resolve the influence of other mechanisms in this regime. Consequently, for measurements at 0.09~T and 2~T, the echo decay is primarily governed by the TLS term, making it difficult to extract reliable values for \( \Gamma_{\mathrm{SD}} \) and \( R_{\mathrm{SD}} \) at elevated temperatures as can be seen in Fig.~\ref{fig:Gamma_SD} and Fig.~\ref{fig:R_SD}. In contrast, at 0~T, the spectral diffusion amplitude \( \Gamma_{\mathrm{SD}} \) remains sufficiently large to contribute measurably even in the presence of strong TLS broadening, allowing for extraction of parameters across the full temperature range. This behavior is qualitatively consistent with trends observed in erbium-doped crystals, where \( \Gamma_{\mathrm{SD}} \) follows a field- and temperature-dependent form such as \( \Gamma_{\mathrm{SD}}(B, T) \propto \text{sech}^2(g \mu_B B / 2kT) \)~\cite{bottger2006optical}, suggesting suppressed spin flip-flop interactions at high fields and low temperatures. While we do not claim to verify this exact form in our fiber system, the relatively high \( \Gamma_{\mathrm{SD}} \) observed at $0 \, \mathrm{T}$ is consistent with this trend.

At the lowest temperatures, we observe that \( \Gamma_{\mathrm{SD}} \) and \( R_{\mathrm{SD}} \) become suppressed, suggesting a reduced influence from spin flip-flop processes. The behavior of \( R_{\mathrm{SD}} \) shown in Fig.~\ref{fig:R_SD} is consistent with our previous study on spectral hole burning in erbium-doped fibers~\cite{bornadel2025hole}, where long-lived Zeeman sublevels and suppressed spin dynamics were observed at millikelvin temperatures.

Finally, Fig.~\ref{fig:Branching_Ratio} shows a gradual variation of the branching ratio with magnetic field, indicating changes in decay pathways between the excited state and ground-state Zeeman levels. We observe an intermediate field region where the transition behavior appears mixed. A more detailed interpretation of this trend is presented in the discussion section.

\section{Discussion and Conclusion\\}\label{ssec:discussion_outlook}

In this study, we explored the optical coherence properties of Er-doped silica fibers (EDFs), we conducted two pulse photon echo (2PPE) experiments at temperatures ranging from $\sim 7$ to $550 \, \mathrm{mK}$ and magnetic fields from $0$ to $2 \, \mathrm{T}$. To investigate the spectral diffusion and decoherence processes, we conducted three pulse photon echo (3PPE) for $0$, $0.09$, and $2 \, \mathrm{T}$ and from $\sim 7$ to $450 \, \mathrm{mK}$.

Using 2PPE measurements, we obtained an effective homogeneous linewidth ($\Gamma_{\text{eff}}$) of approximately $8 \, \mathrm{kHz}$ at $\sim 7 \, \mathrm{mK}$ and $0.09 \, \mathrm{T}$, making this magnetic field optimal with respect to minimizing $\Gamma_{\text{eff}}$ based on its magnetic-field dependence. However, while $0.09 \, \mathrm{T}$ yields the narrowest linewidth, it is not optimal in terms of the zero-delay echo intensity, $I_0$: increasing the field from $0.09$ to $2 \, \mathrm{T}$ broadens $\Gamma_{\text{eff}}$ from $\sim 8$ to $\sim 12 \, \mathrm{kHz}$ but raises $I_0$ from $\sim 30 $ to $\sim 80 \, \mathrm{\%}$, which may represent a more favorable operating point when considering both linewidth and $I_0$ together.

To further explore the observed temperature and magnetic field dependence of the $\Gamma_{\text{eff}}$ that we saw in section \ref{sectionA2} one possible interpretation involves coherently coupled two-level systems (TLSs). Previous studies \cite{burin1994low, ding2020probing} have indicated that coupled TLS pairs can play a significant role in decoherence under specific temperature conditions. Within the standard TLS framework, the density of states is given by \( \rho(E) \propto E^{0.3} \), where \(E\) is the energy of the TLS \cite{huber1984low, ding2020probing}. Using this expression, the temperature dependence of the homogeneous linewidth can be approximated as \( \Gamma_h(T) \propto T^{1.3} \). At sufficiently low temperatures, when phonon scattering becomes weaker than the coupling strength within TLS pairs \cite{burin1994low, ding2020probing}, coherently coupled TLS pairs are formed. For these coupled TLSs, the density of states is expressed as \( \rho(E^\prime) \propto T {E^\prime}^{-2} \), where $E^\prime$ is the energy of the pair of TLSs, leading to \( \Gamma_h(T) \propto T^0 \). This indicates that below a certain temperature, the homogeneous linewidth becomes constant. As shown in Fig.~\ref{fig:Temp_dependence}, this transition in the temperature dependence is clearly observed in our measurement of $\Gamma_{\text{eff}}$.

We also observed an additional magnetic field–dependent term in $\Gamma_{\text{eff}}$ at $T \approx 7 \, \mathrm{mK}$, as shown in Fig.~\ref{fig:Field_dependence} and Eq.~(\ref{eq:1.3}). This behavior may suggest that, at sufficiently low temperatures, a different class of TLS modes becomes active in the EDF sample. The term $\alpha_1 \left(1 - \exp\left(-\frac{g_1 \mu_B B}{k_B T}\right)\right)$ appears to impose a saturation limit on $\Gamma_{\text{eff}}$ at higher magnetic fields. Our fitting values implies that for this newly active class of TLSs, the effective $g$-factors for the coupled Er-magnetic TLS systems is significantly smaller than that of the dominant decoherence source observed at higher fields ($g_1 \ll g_2$).

3PPE measurements provided insight into spectral diffusion processes and long-timescale decoherence in EDF. By fitting the decay of echo intensity as a function of waiting time \( t_{23} \), we separated the contributions from intrinsic homogeneous broadening, fast spectral diffusion due to spin flip-flops, and slow logarithmic broadening attributed to TLS. The extracted dephasing parameters revealed reduced TLS coupling and spin dynamics at low temperatures and high magnetic fields, consistent with long spin lifetimes observed in our spectral hole burning study~\cite{bornadel2025hole}. As shown in Fig.~\ref{fig:Gamma_TLS}, the TLS coupling strength decreases by over an order of magnitude below 100~mK, becoming negligible at millikelvin temperatures. 

Fig.~\ref{fig:Branching_Ratio} shows the magnetic field dependence of the branching ratio from the excited state to the intermediate Zeeman level. We observed reduced branching at both low and high fields, while the intermediate field region around 0.09~T exhibits higher values. This behavior may reflect a crossover regime: at low fields, hyperfine splitting governs selection rules, while at high fields, Zeeman splitting dominates. In between, both mechanisms may be active, relaxing the selection rules and allowing increased decay into the other Zeeman sublevel. A complete microscopic explanation of these branching trends is beyond the scope of this study.

We observed an improvement of more than two orders of magnitude in the effective homogeneous linewidth, $\Gamma_{\text{eff}}$, of the EDF at $\sim 7 \, \mathrm{mK}$ compared to previous results at $T = 700 , \mathrm{mK}$ \cite{veissier2016optical}. Further reduction in $\Gamma_{\text{eff}}$ may be achieved by replacing co-dopant ions in EDF that have non-zero nuclear spin with those that possess zero nuclear spin such as $^{27}$Al \cite{wannemacher1991nuclear,staudt2006investigations}, thereby suppressing spin dynamics in the environment. Another promising approach is to use isotopically purified EDFs containing either only $^{167}$Er ions (with nuclear spin) or isotopes with zero nuclear spin, to further enhance optical coherence. It is also possible that the measured $\Gamma_{\text{eff}} \approx 8 \, \mathrm{kHz}$ contains a contribution from excitation-induced decoherence processes, such as instantaneous spectral diffusion (ISD) \cite{graf1998photon}
To investigate this, $\Gamma_{\text{eff}}$ can be measured as a function of the optical pulse power in the 2PPE sequence, enabling the determination of the power-independent linewidth by extrapolating to zero excitation power, following the model described in \cite{thiel2014measuring}. The ISD can be further probed by introducing a detuned scrambler optical pulse into the 2PPE sequence and monitoring its effect on both the echo intensity and $\Gamma_{\text{eff}}$ \cite{graf1998photon}.

The combination of this narrow $\Gamma_{\text{eff}}$ and the long spin lifetime reported in \cite{bornadel2025hole} highlights the potential of EDFs as a platform for quantum memories in future quantum networks.

\section*{Acknowledgments}

The authors thank Dr. Charles W. Thiel and Faezeh Kimiaee Asadi for valuable discussions. This work was supported by the Government of Alberta's Major Innovation Fund Project on Quantum Technologies, the Canadian Foundation for Innovation Infrastructure Fund (CFI-IF), the Natural Sciences and Engineering Research Council of Canada (NSERC) through the Alliance Quantum Consortia Grants QUINT and ARAQNE, and by the National Research Council of Canada (NRC) through the High Throughput Secure Network Challenge Program.

\bibliographystyle{apsrev4-1}
\bibliography{main}   

\end{document}